\documentclass[acus]{JAC2003}

\usepackage{graphicx}

\begin{document}
\title{eRHIC DETECTOR DESIGN STUDIES - IMPLICATIONS AND CONSTRAINTS ON THE ep(A) INTERACTION-REGION DESIGN}

\author{B. Surrow\thanks{surrow@mit.edu}, Massachusetts Institute of Technology, Cambridge, MA 02139, USA \\
A. Deshpande, Stony Brook University, Stony Brook, NY 11974, USA\\
J. Pasukonis, Massachusetts Institute of Technology, Cambridge, MA 02139, USA }

\maketitle

\begin{abstract}
 An electron-proton/ion collider facility (eRHIC) is under consideration at Brookhaven
 National Laboratory (BNL).  Such a new facility will require the design and construction of a new optimized detector
 profiting from the world's first ep collider facility HERA at DESY.
 The detailed design is closely coupled to the design of the interaction region and thus to the
 machine development work in general. Implications and constraints on the ep(A) interaction-region design
 will be discussed.
\end{abstract}

\section{Introduction}

 The high energy, high intensity polarized electron/positron beam
 ($5-10\,$GeV/$10\,$GeV) facility (eRHIC) to collide with the existing RHIC heavy ion ($100\,$GeV per nucleon) and
 polarized proton beam ($50-250\,$GeV) would
 significantly enhance the exploration of fundamental aspects of Quantum Chromodynamics
 (QCD), the underlying quantum field theory of strong interactions.
 A detailed report on the accelerator and interaction region design of this new collider facility
 has been completed based on studies performed jointly by BNL and MIT-Bates in collaboration with
 BINP and DESY \cite{1-ref}.

\section{Outline of the \lowercase{e}RHIC detector design}

The following discussion will be restricted to the nominal eRHIC collider mode operation of a $10\,$GeV electron/positron
beam colliding with a $250\,$GeV proton beam. Simple four vector kinematics in e-p collisions which involves an electron and
a proton in the initial state and a scattered electron along with a hadronic final state in the final state can be used to study
analytically the energy and angular acceptance of the scattered electron and hadronic final state as a function of the main kinematic
quantities in deep-inelastic scattering (DIS), $x$ and $Q^{2}$ \cite{3-ref}. This provides a first understanding of the final state
topology. $Q^{2}$ is the negative four momentum transfer squared between the incoming and scattered electron. The Bjorken scaling variable
$x$ is interpreted in the Quark-Parton model as the fraction of the proton momentum carried by the struck quark. The hadronic final state consists
of the current jet which emerges from the struck quark characterized by its polar angle and energy.

\begin{figure}[ht]
\centering
\includegraphics*[width=85mm]{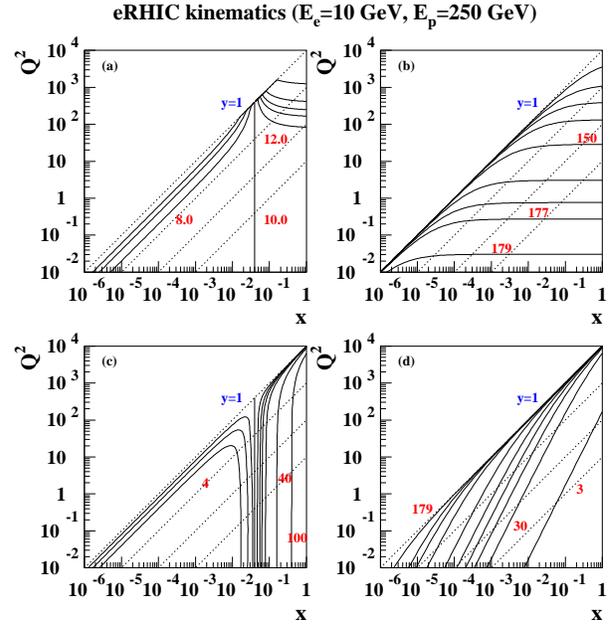}
\caption{The dashed line represents lines of constant $y$ values (1, 0.1, 0.01, 0.001). The electron beam energy amounts to $10\,$GeV whereas the proton beam energy is $250\,$GeV. Isolines of constant electron energies (4, 6, 8, 10, 12, 14, 16, 20 and 40GeV) (a), electron scattering angles ($30^{\circ}$ - $150^{\circ}$ in steps of $30^{\circ}$ and $170^{\circ}$, $175^{\circ}$, $177^{\circ}$ and $179^{\circ}$) (b), current jet energies (4, 6, 8, 12, 14, 16, 20, 40 and 100GeV) (c) and current jet angles ($3^{\circ}$, $30^{\circ}$-$150^{\circ}$ in steps of $30^{\circ}$ and $170^{\circ}$, $175^{\circ}$, $177^{\circ}$ and $179^{\circ}$) (d).}
\label{fig1}
\end{figure}

Figure \ref{fig1} shows isolines of constant electron energy (a) and scattering angle (b) as well as lines of constant $y$
values (1, 0.1, 0.01, 0.001). The kinematic variable $y$ is given in terms of $x$ and $Q^{2}$ ($y=Q^{2}/(sx)$) and refers to the inelasticity in the rest
frame of the proton. The kinematic limit is given by $y=1$. The scattering angle is measured with respect to the incoming proton beam which defines
the positive $z$ axis. Electron tagging acceptance down to at least $177^{\circ}$ will be necessary to provide acceptance in $Q^{2}$ down to $0.1\,$GeV$^{2}$.
The energy of the scattered electron is less than $10\,$GeV for a large fraction of the kinematic region and is in particular small in the region of
low $x$ and medium to low values in $Q^{2}$. This sets stringent requirements on trigger and reconstruction efficienies.
Figure \ref{fig1} shows isolines of constant current jet energy (c) and scattering angle (d). The energy of the current jet is rather small in the low
$x$ and medium to low $Q^{2}$ region and overlaps to some extend with the scattered electron. This will require e/h separation in particular in the rear direction, i.e.
the direction of larger electron scattering angles with respect to the incoming proton beam. The current jet energy increases towards the forward
direction in the region of high $x$ and $Q^{2}$ values. The capability to measure larger jet energies will be
in partiuclar important in the forward direction, i.e. the incoming proton beam direction.

The following minimal requirements on a future eRHIC detector can be made:

\begin{Itemize}
\item Measure precisely the energy and angle of the scattered electron (Kinematics of DIS reaction)
\item Measure hadronic final state (Kinematics of DIS reaction, jet studies, flavor tagging, fragmentation studies, particle ID)
\item Missing transvere energy measurement (Events involving neutrinos in the final state, electro-weak physics)
\end{Itemize}

In addition to those demands on a central detector, the following
forward and rear detector systems are crucial:

\begin{Itemize}
\item Zero-degree photon detector to control radiative corrections and measure Bremsstrahlung photons for luminosity measurements
\item Tag electrons under small angles (Study of the non-perturbative/perturbative QCD transition region and luminosity measurements from
Bremsstrahlung ep events)
\item Tagging of forward particles (Diffraction and nuclear fragments)
\end{Itemize}

\begin{figure}[!h]
\centering
\includegraphics*[width=80mm]{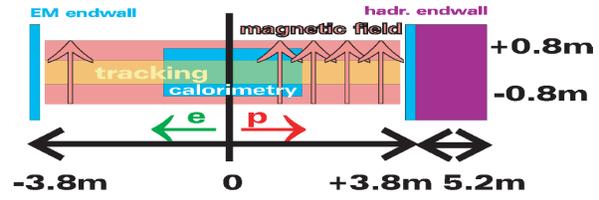}
\caption{Conceptual detector layout focusing on forward physics with
a $7\,$m long dipole field and an interaction region without machine
elements extending from $-3.8\,$m to $+5.8\,$m.} \label{fig3}
\end{figure}

Optimizing all the above requirements is a challenging task. Two
detector concepts have been considered so far. One, which focuses on
the forward acceptance and thus on low-$x$/high-$x$ physics which
emerges out of HERA-III detector studies \cite{5-ref}. This detector
concept is based on a compact system of tracking and central
electromagnetic calorimetry inside a magnetic dipole field and
calorimetric end-walls outside. Forward produced charged particles
are bent into the detector volume which extends the rapidity
coverage compared to existing detectors. A side view of the detector
arrangment is shown in Figure \ref{fig3}. The required machine
element-free region amounts to roughly $\pm 5$m. This clearly limits
the achieveable luminosity in a ring-ring configuration.

\begin{figure}[hbt]
\centering
\includegraphics*[width=80mm]{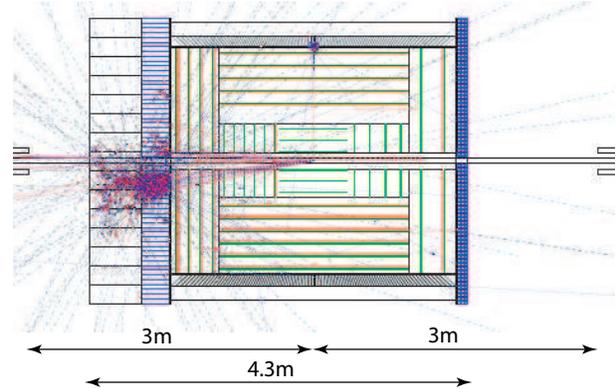}
\caption{Side view of the GEANT dector implementation as part of the
ELECTRA simulation and reconstruction package \cite{4-ref}. A
deep-inelastic scattering event resulting from a LEPTO simulation is
overlayed with $Q^{2}=361\,$GeV$^{2}$ and $x=0.45$.} \label{fig2}
\end{figure}

The second design effort focuses on a wide acceptance detector
system similar to the current HERA collider experiments H1 and ZEUS
to allow for the maximum possible $Q^{2}$ range. The physics program
demands high luminosity and thus focusing machine elements in a
ring-ring configuration which are as close as possible to the
interaction region while preserving good central detector
acceptance. This will be discussed in more detail in the next
section. A simulation and reconstruction package called ELECTRA has
been developed to design a new eRHIC detector at BNL \cite{4-ref}.
Figure~\ref{fig2} shows a side view of a GEANT detector
implementation of the above requirements on a central detector. The
hermetic inner and outer tracking system including the
electromagnetic section of the barrel calorimeter is surrounded by
an axial magnetic field. The forward calorimeter is subdivided into
hadronic and electromagnetic sections based on a conventional
lead-scintillator type. The rear and barrel electromagnetic consists
of segmented towers, e.g. a tungsten-silicon type. This would allow
a fairly compact configuration. Other options based on a crystal
rear and barrel electromagnetic calorimeter are under study. The
inner most double functioning dipole and quadrupole magnets are
located at a distance of $\pm 3$m from the interaction region. An
initial interaction region design assumed those inner most machine
elements at $\pm 1$m. This would significantly impact the detector
acceptance and is therefore not considered. More details on the
interaction region design can be found in \cite{1-ref}.

The bunch crossing frequency amounts to roughly $30\,$MHz. This sets
stringent requirements on the high-rate capabililty of the tracking
system. This makes a silicon-type detector for the inner tracking
system (forward and rear silicon disks together with several silicon
barrel layers) together with several GEM-type outer tracking layers
a potential choice. The forward and rear detector systems have not
been considered so far. The design and location of those detector
systems has to be worked out in close collaboration to accelerator
physicists since machine magnets will be potentially employed as
sepectrometer magnetes and thus determine the actual detector
acceptance and ultimately the final location. It is understood that
demands on optimizing the rear/forward detector acceptance might
have consequences on the machine layout and is therefore an
iterative process.

\section{Considerations on the detector/machine interface}

The following section provides an overview of some aspects of the detector/machine interface. The specification of those items has only recently been started:

\subsection{Synchrotron radiation}

The direct synchrotron radiation has to pass through the entire interaction region
before hitting a rear absorber system. This requires that the geometry of the beam pipe is designed appropriately with changing shape along the longitudinal beam
direction which includes besides a simulation of the mechanical stress also the simulation of a cooling system of the inner beam pipe. The beam pipe design has
to include in addidition the requirement to maximize the detector acceptance in the rear and forward direction. Furthermore the amount of dead material has
to be minimized in particular to limit multiple scattering (track reconstruction) and energy loss for particles under shallow angles (energy reconstruction). The distribution
of backscattered synchrotron radiation into the acutal detector volume has to be carefully evaluated. An installation of a collimator system has to be worked out. Those
items have been started in close contact to previous experience at HERA \cite{6-ref}.

\subsection{Location of inner machine elements}

The demand of a high luminosity ep/eA collider facility requires the installaton of focusing machine elements
as close as possible to the central detector. An interaction region design with machine elements as close $\pm 1\,$m to the
interaction region which has been presented in \cite{1-ref} would significantly limit the achievable detector acceptance. A new
scheme has been presented in \cite{2-ref} which provides a machine-element free region of $\pm 3\,$m at the expense of approximatley
half the luminosity for the interaction region design presented in \cite{1-ref}. A linac-ring option would not be limited by beam-beam
effects compared to a ring-ring configuration. Even larger luminosities could be achieved with a machine-element free region
of approximatley $\pm 5\,$m. This scheme has been presented in \cite{7-ref}.

\subsection{Rear tagging system}

The need for acceptance of scattered electrons beyond the central detector acceptance is driven by the need for luminosity
measurements through ep/eA Bremstrahlung and photo-production physics. Besides that a calorimeter setup to tag radiated photons
from inital-state radiation and Bremsstrahlung will be necessary. The scattered electrons will pass through the machine elements
and leave the beam pipe through special exist windows. The simulation of various small-angle calorimeter setups has been started.
This will require a close collaboration with the eRHIC machine design efforts to aim for an optimal detector setup.

\subsection{Forward tagging system}

The forward tagging system beyond the central detector will play a crucial role in diffractive ep/eA physics.
The design of a forward tagger system based on forward calorimetry and Roman pot stations is foreseen. Charged particles will be deflected
by forward machine elements. This effort will require as well a close collaboration with the eRHIC machine design efforts to ensure the best
possible forward detector acceptance.


\end{document}